\documentclass[lettersize,journal,twoside]{IEEEtran}
\usepackage{amsmath,amsfonts}
\usepackage{algorithmic}
\usepackage{algorithm}
\usepackage{array}
\usepackage[caption=false,font=normalsize,labelfont=sf,textfont=sf]{subfig}
\usepackage{textcomp}
\usepackage{stfloats}
\usepackage{url}
\usepackage{verbatim}
\usepackage{graphicx}
\usepackage{cite}
\usepackage{booktabs}
\usepackage{eqparbox}
\usepackage{xcolor}
\usepackage[]{hyperref}
\begin{document}

\title{RADS-Checker: Measuring Compliance with Right of Access by the Data Subject in Android Markets}

\author{Zhenhua Li, Zhanpeng Liang, Congcong Yao, Jingyu Hua, and Sheng Zhong
\thanks{The first two authors contributed equally to this work.}
}



\maketitle

\begin{abstract}
The latest data protection regulations worldwide, such as the General Data Protection Regulation (GDPR), have established the Right of Access by the Data Subject (RADS), granting users the right to access and obtain a copy of their personal data from the data controllers. This clause can effectively compel data controllers to handle user personal data more cautiously, which is of significant importance for protecting user privacy. However, currently there is no research that systematically examines whether RADS has been effectively implemented in mobile apps, which are the most common personal data controllers. In this study, we propose a compliance measurement framework for RADS in apps. For a specific app, we first analyze its privacy policy text using NLP techniques such as GPT-4 to verify whether it clearly declares that it offers RADS to users and provides specific details on how the right can be exercised. Next, we assess the authenticity and usability of the identified implementation methods by submitting data access requests to the app. Finally, for the obtained data copies, we further verify their completeness by comparing them with the personal data actually collected by the app during run-time, as captured by Frida Hook. We analyze a total of 1,631 apps in the European app market G and the Chinese app market H. The results show that less than 54. 50\% and 37. 05\% of the G and H apps, respectively, explicitly state in their privacy policies that they can provide users with copies of their personal data. Furthermore, in both mobile apps markets, less than 20\% of the apps could truly provide users with their data copies. Finally, among the data copies obtained, only about 2. 94\% from G pass the completeness verification.
\end{abstract}


\section{Introduction}
\IEEEPARstart{W}{ith} smartphones playing an increasingly important role in daily life, concerns about user privacy related to mobile app usage have garnered significant attention from society and governments worldwide. To protect user privacy, various data protection laws and regulations have been enacted worldwide, such as the General Data Protection Regulation \cite{gdpr2016} (GDPR) and the Personal Information Protection Law of the People's Republic of China \cite{pipl2021} (PIPL). Without exception, these laws contain a specific provision (Art. 15 of GDPR and Art. 45 of PIPL) granting users a core right: the Right of Access by the Data Subject (RADS). This right entitles users to access their personal data and obtain a copy from the data controller, as specified in Art. 15(3) of GDPR: ``The controller shall provide a copy of the personal data undergoing processing". Personal data means any information relating to an identified or identifiable natural person (‘data subject’) (in Art. 4(1) of GDPR). By exercising RADS, users can more directly assess the privacy risks associated with using a particular app. Moreover, this practice can also enhance data processing transparency, prompting data controllers to handle user data more cautiously and ensure compliance with ethical and legal requirements. Thus, RADS is crucial in safeguarding user privacy.

Unfortunately, despite the clear legal stipulations regarding RADS, through manual inspection of various popular apps, we have discovered that many of them have not fully implemented RADS, with serious issues like failing to provide data copies and not responding to user requests. To comprehensively analyze the compliance situation of RADS, this paper proposes a compliance measurement framework for RADS in apps: RADS-Checker, and conducts large-scale compliance measurements and comparisons across two major Android markets in the 
Europe and China (anonymously as G and H). For a specific app, we mainly try to answer three questions: \textbf{(1) Does the app declare that it provides RADS and specify the detailed implementation methods? (2) Can users view their personal data and obtain a copy through the implementation methods informed by the app, and how usable are these methods? (3) Is the personal data copy obtained from the app complete?}


Typically, apps elaborate on the various privacy rights available to users in their privacy policies, including RADS. Therefore, we analyze the texts of privacy policies of apps to answer the first question. Specifically, according to our manual analysis of the privacy policies of tens of popular apps, we find that declarations regarding RADS can be generally divided into two categories as the examples shown in Figure \ref{fig:Examples of RADS statements in privacy policy(from Shopee and Klook)}: one clearly states that users can request a copy of their personal data, called Data Copy Access Right (DCAR). The other merely makes vague claims that users can access personal information, but does not explicitly state to provide copies of personal data, called Vague Data Access Right (VDAR). Similarly, the implementation methods can be divided into three categories: Email Contact, Account Settings, and Webform Submission. Based on these classifications, we first use NLP technology to automatically extract paragraphs related to RADS and its implementation methods from privacy policies, and then use the large language model GPT-4 \cite{openai2024gpt4} to identify RADS declarations and implementation methods from the extracted paragraphs.

For the second question, after we identify the implementation methods, we mainly focus on the practicality of these methods, including two aspects: \textit{authenticity}, i.e., whether users can truly view their personal information or obtain personal data copies through these methods, and \textit{usability}, i.e., what is the user experience like during the exercise of RADS? We conduct practicality experiments on 200 apps from each of the two app markets. Specifically, we exercise the right according to the implementation methods mentioned in each app's privacy policy, for example, we can operate by selecting the ``access personal information" or ``obtain data copies" options in the app settings interface for the account settings method. Subsequently, we evaluate the authenticity based on the feedback received. Regarding the user experience, we consider two dimensions: feedback duration and the UI depth of settings.

For the third question, we primarily verify whether the personal data copy obtained is complete, which means it should include all personal data actually collected by the app. For this, we select ten categories of sensitive information related to user privacy and identify 26 Android native APIs related to these sensitive data categories. Subsequently, we use the Frida-based \cite{frida} dynamic analysis framework to dynamically monitor the apps and collect the sensitive data collected by the apps during runtime. Finally, we compare these data against the data copy provided by the app to verify the data copy completeness.

In summary, we systematically analyze the compliance of RADS with the apps in mainstream markets to reveal the gap between legal stipulations and app practices. The main contributions and findings of this paper are as follows:

\textbf{Identification of RADS declarations and implementation methods}: This study employs natural language processing technology and the large language model GPT-4 to automatically identify RADS declarations and implementation methods in a privacy policy. We analyze 600 apps in G and 1031 apps in H, finding that less than 54.50\% and 37.05\% of apps in G and H, respectively, fully comply with declaring the right to access data copies in their privacy policies. Furthermore, approximately 32\% of apps in both markets do not declare RADS. Lastly, among the apps that provide RADS, about 64.22\% provide the email contact method, 68.21\% provide the account settings method, a few provide the webform submission method, and about 10.33\% do so.

\textbf{Practicality assessment of the implementation methods}: Regarding the practicality of implementation methods, this study comprehensively assesses the authenticity and usability in 200 apps from each app market. We find that less than 20\% apps in both markets could actually provide users with data copies, with 17\% in G and about 19\% in H. Additionally, among the three methods of implementation, about 20.34\% of apps offering account settings could actually provide data copies, while the offering email contact method and the webform submission method have corresponding proportions of 13.33\% and 28.57\%. Lastly, from the usability perspective, the account settings method typically provides feedback within one day, while the other two methods perform poorly. 

\textbf{Completeness verification of the data copy}: Using the Frida-based dynamic data monitoring framework, this study for the first time conducts an empirical analysis of the completeness of personal data copies provided by apps. We find that in our survey of a total of 400 apps, all apps in H fail to ensure the completeness of data copies, while in G, only 2.94\% of apps show no missing issues.

\section{Background And Problem Definition}
\subsection{Background}\label{chap:two_first}
\textbf{Definition of Personal Information:} In legal contexts, the definition of personal information is quite broad, including various types of information related to the identifiable natural person. For instance, GDPR defines personal information as follows:“‘personal data’ means any information relating to an identified or identifiable natural person (‘data subject’).” Similarly, PIPL defines personal information as “various kinds of information related to identified or identifiable natural persons recorded by electronic or other means, excluding information processed anonymously.”

Besides, various countries have also provided definitions of personal information. For example, the European Commission defines personal information as “any information that relates to an identified or identifiable living individual.”\cite{eu_personal_data}, and gives examples of personal information: Location data; An Internet Protocol (IP) address; The advertising identifier of your phone. Furthermore, the China National Standardization Administration (CNSA)\cite{cn_personal_data} has issued the ``Information security technology—Personal information security specification," which explicitly states that personal information refers to various information recorded by electronic or other means that can identify a specific natural person or reflect a specific natural person's activities, either independently or in combination with other information. Examples include: Network identity identification information; Personal frequently used device information; Personal location information.

\begin{figure}[h]
  \centering
  \includegraphics[width=\linewidth]{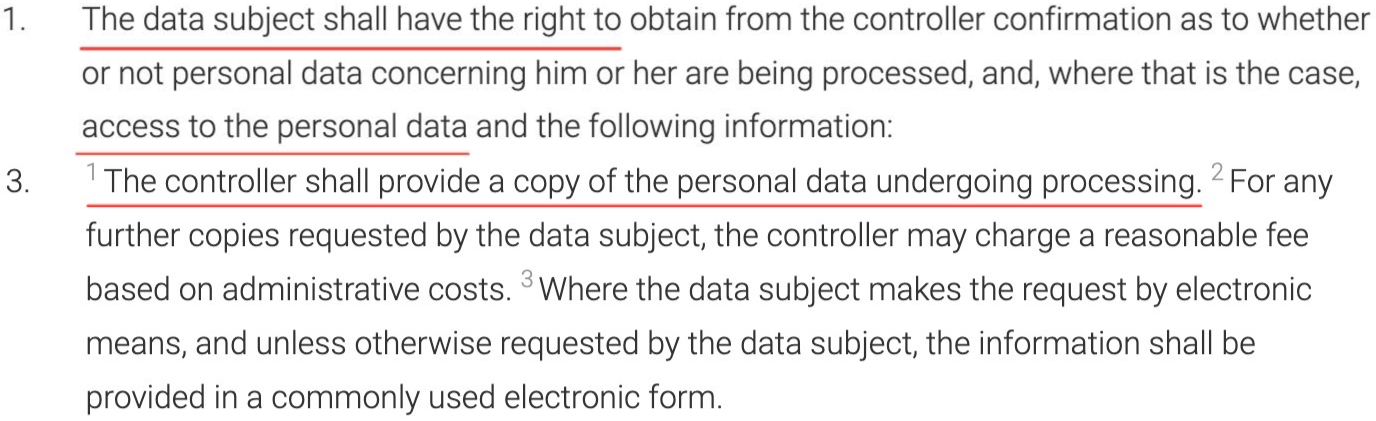}
  \caption{Provisions regarding RADS under the GDPR (Art. 15)}
  \label{fig:Provisions regarding RADS under the GDPR}
\end{figure}
\begin{figure}[h]
  \centering
  \includegraphics[width=\linewidth]{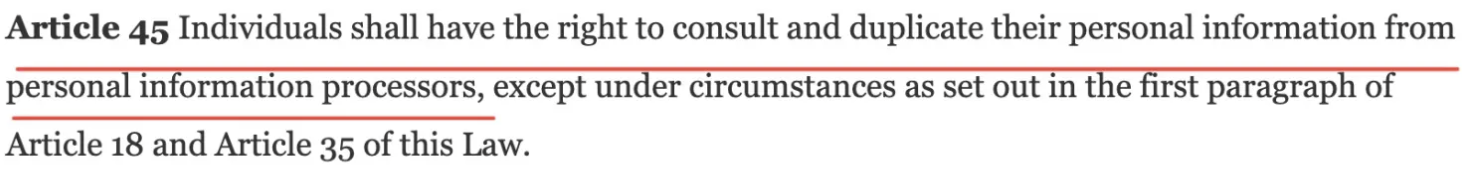}
  \caption{Provisions regarding RADS under the PIPL (Art. 45)}
  \label{fig:Provisions regarding RADS under the PIPL}
\end{figure}

\begin{figure*}[h]
  \centering
  \includegraphics[width=0.7\linewidth]{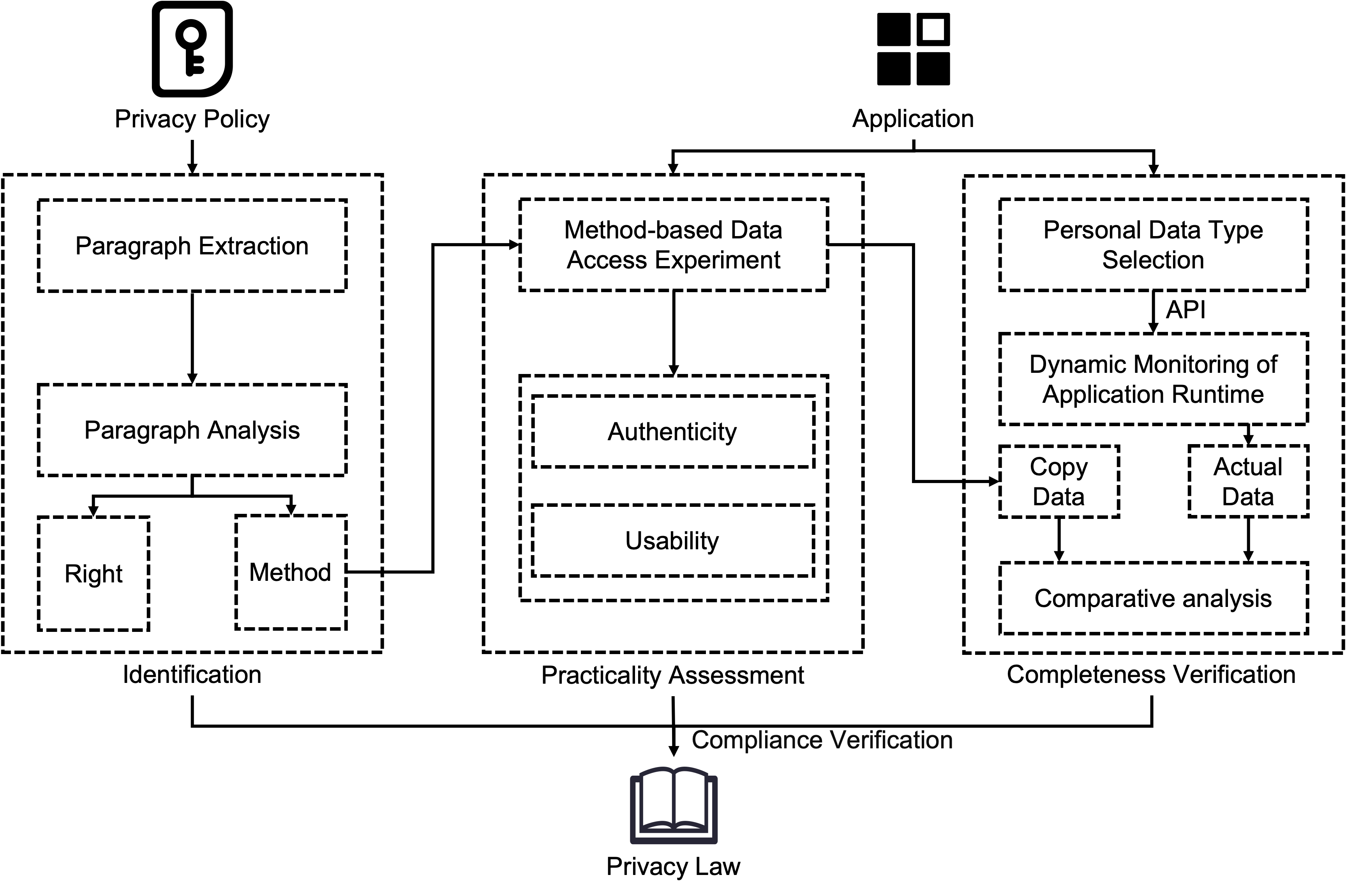}
  \caption{The High-level Architecture of the RADS-Checker Framework}
  \label{fig:The High-level Architecture of the RADS-Checker Framework}
\end{figure*}

\textbf{Right of Access by the Data Subject (RADS):} As personal data becomes increasingly valued by society, more and more data protection laws have been enacted in many regions and countries. Among them, the most typical are the GDPR and PIPL. As illustrated in Figure \ref{fig:Provisions regarding RADS under the GDPR} and Figure \ref{fig:Provisions regarding RADS under the PIPL}, both laws explicitly stipulate RADS, i.e., the data subject shall have the right to access to the personal data and be provided a copy of the personal data by the data controller.

\subsection{Problem Definition}\label{chap:two_two}
Our research question involves investigating the compliance of RADS in real-world apps. To conduct a detailed analysis, we mainly analyze three significant questions:

\noindent\textbf{RQ 1: Do mobile apps declare RADS and the detailed implementation methods to users?}

 Privacy policies serve as legal documents through which apps disclose to users about the collection, storage, and use of personal data. They are crucial channels for users to understand their personal data rights within the app. Therefore, our first step is to examine whether the privacy policies provided by the apps declare RADS complying with relevant laws. If the app clearly articulates this fundamental data right to users, has it also explicitly informed them about the implementation methods to exercise RADS?

\noindent\textbf{RQ 2: How practical are the implementation methods declared by the apps?}

We need to further assess the practicality of the implementation methods disclosed in the privacy policies. This includes two aspects: firstly, the authenticity of the methods declared, i.e., whether the methods mentioned can actually allow users to view their personal information or obtain the data copies of themselves. Secondly, the usability of the methods, i.e., what is the experience of users in the process of exercising RADS?

\noindent\textbf{RQ 3: Are the data copies provided by the apps complete?}

 After confirming that users can obtain personal data copies through these methods, we will further explore the issue of the completeness of these data copies, i.e., whether the data copies include all of the user's personal data that the app have actually collected.
 
\section{Methodology}
In this section, we elaborate on a compliance measurement framework for RADS in Android apps: RADS-Checker. As illustrated in Figure \ref{fig:The High-level Architecture of the RADS-Checker Framework}, the pipeline of RADS-Checker includes three parts:

(1) \textbf{Identification of RADS declarations and implementation methods}: Firstly, we conduct automated analysis of the privacy policy text, identifying the declarations related to RADS and the specific methods for exercising RADS.

(2) \textbf{Practicality assessment of the implementation methods}: Secondly, we conduct experiments on data access actions corresponding to the implementation methods identified in the previous step, and then evaluate the authenticity and usability of the implementation methods based on the users' experience and feedback results.

(3) \textbf{Completeness verification of the data copy}: Finally, we perform dynamic monitoring of the app based on APIs related to sensitive personal information, and compare the actual data collected with the data copy obtained from the app, to verify the completeness of the data copy provided by the app.

\subsection{Identification of RADS Declarations and Implementation Methods}\label{chap:three_first}

\begin{figure}[h!]
  \centering
  \includegraphics[width=\linewidth]{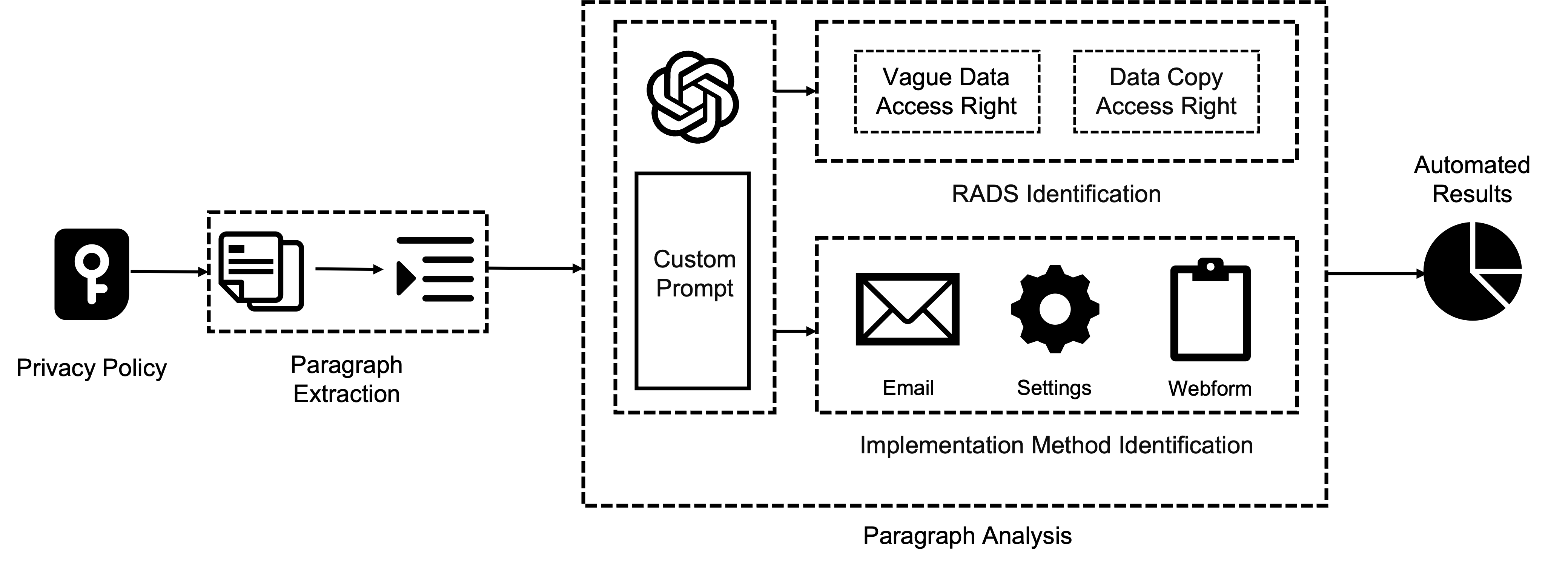}
  \caption{The Pipeline for Identifying RADS Declarations and Implementation Methods}
  \label{fig:The process for identifying RADS and implementation methods}
\end{figure}

In this part, we identify the RADS declarations and corresponding implementation methods from the privacy policy text, focusing primarily on two issues: firstly, whether the app explicitly declares the RADS to users; secondly, what methods the app offers for users to exercise RADS. Privacy policies, serving as declarations of privacy-related terms for apps, are crucial for users to understand how their personal information is processed within the apps. Therefore, as shown in Figure \ref{fig:The process for identifying RADS and implementation methods}, using natural language processing technology, we first extract paragraphs related to RADS and the corresponding implementation methods from privacy policies. Subsequently, based on the large language model GPT-4, we complete the identification and classification of RADS and methods in the Android app market.

\textbf{Data Preparation and Preprocessing}: Initially, we gather texts of privacy policies from applications listed on platforms G and H. On the official market website, we navigate to the interface corresponding to each application utilizing XPath (XML Path Language). Given the generally consistent layout of application interfaces, identifying the XPath for the privacy policy link within the app interface allows us to directly access the link to the app's privacy policy page, and subsequently proceed to enter the app's privacy policy. To streamline this process, we utilize the \textit{Selenium Webdriver} \cite{selenium} to simulate user interactions, thereby automating the retrieval of HTML source code from these pages. Subsequently, we employ \textit{HtmlToPlaintext} \cite{Policylint2019} to convert the HTML source into plain text, extracting the substantive content of the privacy policies.

\begin{table}[t]
\centering
  \caption{The Category of Privacy Policy Paragraphs \cite{polisis_classifiers}}
  \label{tab:The categorization of privacy policy paragraphs}
  \begin{tabular}{cc}
    \toprule
    Category\\
    \midrule
    First Party Collection/Use \\
    Third Party Sharing/Collection \\
    User Access, Edit and Deletion \\
    Data Retention \\
    Data Security \\
    International and Specific Audiences \\
    Do Not Track \\
    Policy Change \\
    User Choice/Control \\
    Introductory/Generic \\
    Practice Not Covered \\
    Privacy Contact Information \\
  \bottomrule
\end{tabular}
\end{table}

\textbf{Paragraph Extraction}: Given that privacy policy texts are usually lengthy and most of the content does not directly relate to RADS, we need to extract descriptive information related to RADS and implementation methods from these texts. To effectively perform our extraction work, we use a pre-trained model \cite{polisis_classifiers} for paragraph classification, which refers to the existing automated privacy policy analysis framework Polisis \cite{Polisis2018}, categorizing the privacy policy texts into 12 different categories based on their content characteristics, as shown in Table \ref{tab:The categorization of privacy policy paragraphs}. Declarations related to RADS fall under the ``User Access, Edit, and Deletion" category. Therefore, we first extract the paragraphs belonging to the ``User Access, Edit, and Deletion" category and treat them as directly related to RADS. Additionally, declarations about implementation methods might appear in the ``Privacy Contact Information" category. However, since a large number of paragraphs belong to the ``Privacy Contact Information" category and most are not related to the methods about RADS, we extract the paragraphs from this category that are adjacent to ``User Access, Edit, and Deletion" paragraphs, considering these to be closely related to the methods for RADS. This approach enables us to obtain information from privacy policies concerning RADS and implementation methods.

\begin{figure}[h]
  \centering
  \includegraphics[width=\linewidth]{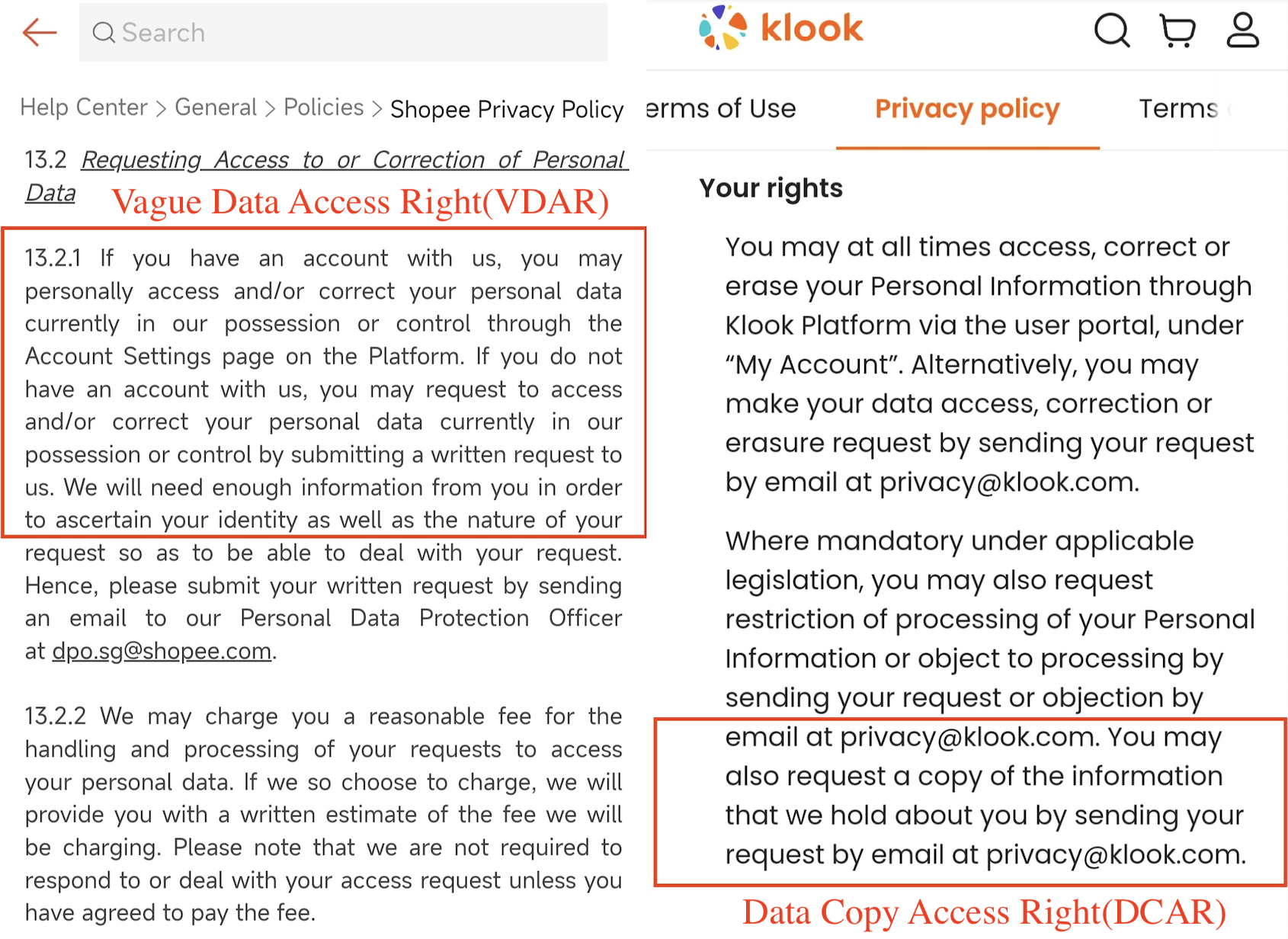}
  \caption{Examples of the RADS Declarations in the Privacy Policy(from Shopee and Klook)}
  \label{fig:Examples of RADS statements in privacy policy(from Shopee and Klook)}
\end{figure}

\textbf{Paragraph Analysis}: Next, we identify whether there are declarations related to RADS in these paragraphs. Based on our analysis of privacy policies in the app market, we find that in most mobile apps' privacy policies, RADS can be divided into two categories, as shown in Figure \ref{fig:Examples of RADS statements in privacy policy(from Shopee and Klook)}: Data Copy Access Right (DCAR) and Vague Data Access Right (VDAR). DCAR makes clear that users have the right to obtain a personal copy of themselves. However, VDAR allows users to view their personal information, but without explicitly stating that a personal data copy can be provided. 

Art.15(3) of GDPR clearly requires data controllers to provide a copy of the personal data under processing. Thus, to comply with legal requirements and ensure that users' privacy and data security are adequately protected, apps' privacy policy should clearly declare DCAR.

In addition to distinguishing the catigories of RADS in the privacy policies, we also categorize the implementation methods about RADS. Based on our manual analysis of privacy policies, these methods can be divided into three main categories: Email Contact, Account Settings, and Webform Submission, with corresponding examples shown in Table \ref{tab:Examples of three types of implementation methods for RADS}.

\begin{table*}
\centering
  \caption{Examples of Implementation Methods for RADS}
  \label{tab:Examples of three types of implementation methods for RADS}
  \begin{tabular}{lp{14cm}}
    \toprule
    Method &Example\\
    \midrule
    Email Contact &Alternatively, you may make your data access, correction or erasure request by sending your request by email at privacy@klook.com.(from Klook) \\
    \midrule
    Account Settings &If you want to see what information we have collected about you, you can request a copy of your data in the Privacy \& Safety section of your User Settings.(from Discord) \\
    \midrule
    Webform Submission &If you want to exercise your right of access or erasure, all you need to do is complete and submit the Data Subject Request for Booking.com Customers form.(from Booking.com) \\
    \bottomrule
  \end{tabular}
\end{table*}

\begin{figure}[h]
  \centering
  \includegraphics[width=\linewidth]{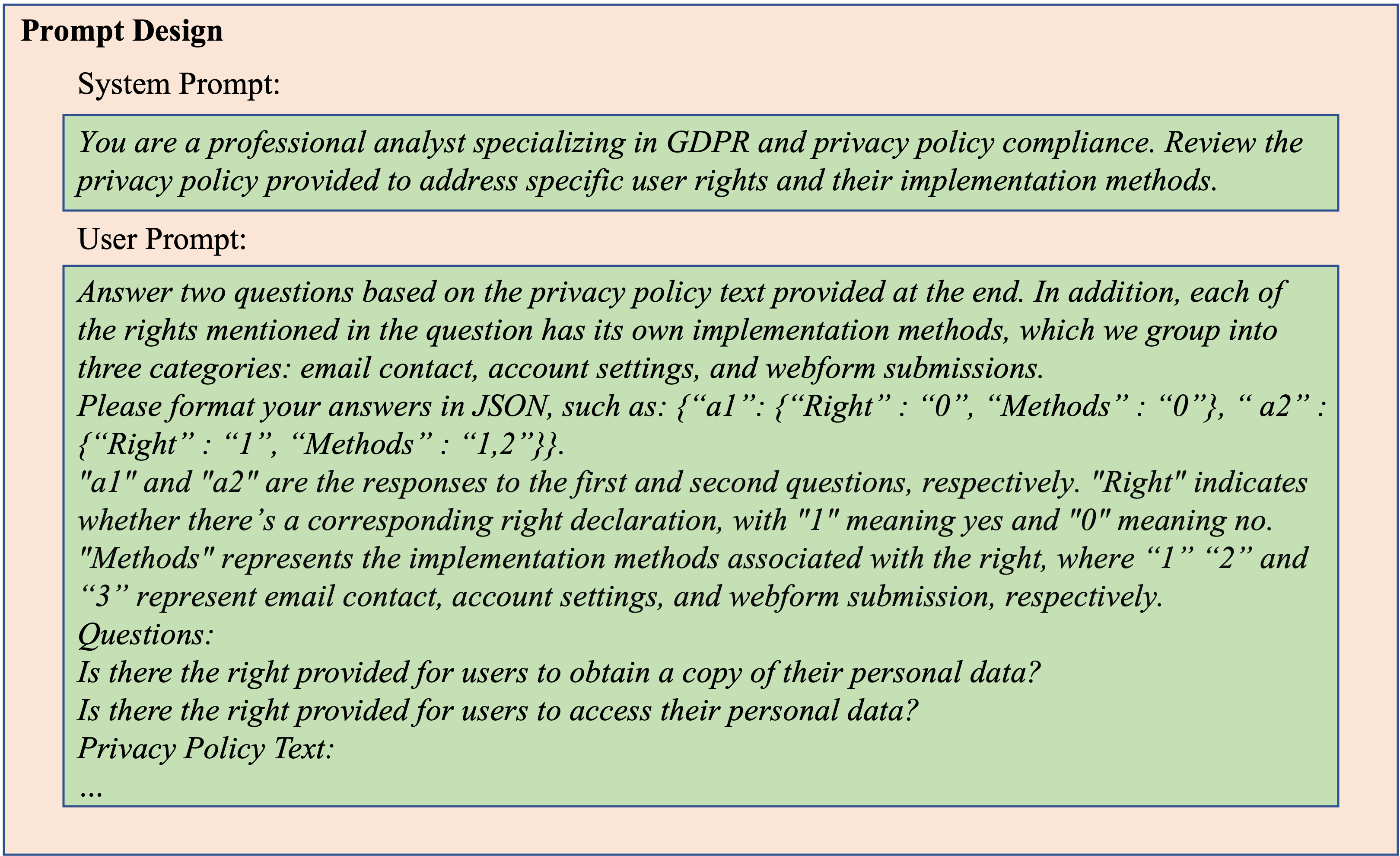}
  \caption{Design of the Prompt for Identifying RADS Declarations and Implementation Methods in Privacy Policy Text}
  \label{fig:Design of the prompt for identifying privacy policy text}
\end{figure}

Next, we use the large language model GPT-4 to automate the identification and classification of RADS and  implementation methods in privacy policies. In the GPT-4 API framework, two core roles are defined: ``System" and ``User". The ``System" role is responsible for defining the model's functional orientation and behavioral guidelines, aiming to guide the model to accurately understand the context and setting of the interaction. In contrast, the ``User" role carries the interactive function, triggering the model to generate the expected responses by asking specific questions and stating requirements. With this in mind, we design a prompt related to the identification and classification of privacy policy texts, as shown in Figure \ref{fig:Design of the prompt for identifying privacy policy text}.

We first set the ``System" to be a professional analyst proficient in GDPR and privacy policies, enabling it to handle our privacy policy texts similarly to a human privacy analyst, and assign it the task of identifying RADS and implementation methods based on the privacy policy. Subsequently, we present our requirements in the ``User", asking two questions to determine whether the provided privacy policy text includes DCAR and VDAR, and to categorize the corresponding methods. The output is formatted in JSON to facilitate subsequent data analysis, with ``a1" and ``a2" representing responses to the first and second questions, respectively. The ``Right" field set to ``1" indicates the presence of a corresponding rights declaration; otherwise, it is ``0". ``Methods" indicates the implementation method corresponding to the right: ``1" for email contact, ``2" for account settings, and ``3" for webform submission. Finally, we attach the privacy policy text at the end of the prompt, completing the full prompt structure. For each application, if the ``Right" in the response to the first question is ``1", it is considered to declare DCAR. If the ``Right" in the response to the first question is ``0" and the ``Right" in the response to the second question is ``1", it is considered to declare VDAR. Otherwise, we consider the application to not declare RADS.

\subsection{Practicality Assessment of Implementation Methods}
After identifying the implementation methods of RADS, we explore the practicality of these methods. We focus on two key aspects: firstly, the authenticity, meaning whether users can actually view their personal data or obtain a data copy through these methods. Secondly, the usability, meaning what is the user's experience during the process? To thoroughly evaluate this aspect, we consider various dimensions, including feedback duration for data access requests and the complexity of the operational steps.

\textbf{Authenticity Assessment}: The authenticity assessment primarily focuses on whether users can truly exercise RADS through the implementation methods provided by the app, including viewing personal information for VDAR and obtaining personal data copies for DCAR.

For each app, we first identify RADS declarations and specific methods by Section \ref{chap:three_first}. Then, depending on the method, we conduct different experiments. For email contact, we send emails to the email address provided in the privacy policy and request to view personal information or obtain a data copy based on the app’s RADS declarations. Similarly, for webform submission, we find the webform link in the privacy policy, fill out the form, and submit it. For account settings, we view personal information through the app's settings or user profile, or find ways to download personal data copies in the privacy settings to exercise RADS.

\textbf{Usability Assessment}: As shown in Table \ref{tab:Evaluation metrics for assessing the usability of the implementation methods}, we design a series of evaluation metrics to quantify usability for different implementation methods.

\begin{table}[h]
  \centering
  \caption{Evaluation Metrics for Assessing the Usability of the Implementation Methods}
  \label{tab:Evaluation metrics for assessing the usability of the implementation methods}
  \begin{tabular}{cc}
    \toprule
    Method &Evaluation Metrics\\
    \midrule
    Email Contact &Feedback Duration\\
    Account Settings &Feedback Duration, UI Depth of Settings\\
    Webform Submission &Feedback Duration\\
  \bottomrule
\end{tabular}
\end{table}

For email contact, we focus on the duration from the user sending a personal information access request via email to receiving a response from the app. This metric is used to quantify the efficiency of email contact in practice. Similarly, for webform submission, feedback duration is also used as the primary consideration.

For account settings, in addition to feedback duration, we also evaluate the depth of data access UI provided in the account settings within the app. This includes the UI depth at which users can view personal information or request a data copy from the initial page within the app. The UI depth of account settings directly affects the complexity of accessing personal information, making it a key metric for assessing usability.

\subsection{Completeness Verification of the Data Copy}\label{chap:three_third}
\begin{figure}[h]
  \centering
  \includegraphics[width=\linewidth]{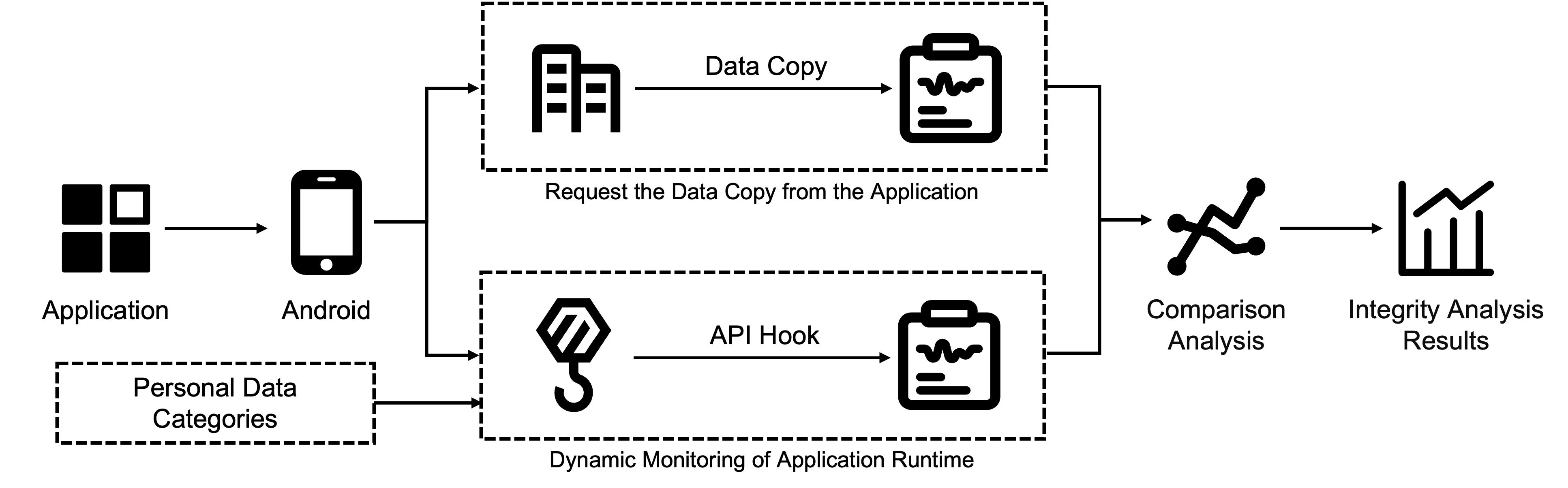}
  \caption{The Pipeline for Completeness Verification of the Data Copy}
  \label{fig:Methodological framework for data completeness verification}
\end{figure}

To verify the completeness of the data copy provided by the app, as shown in Figure \ref{fig:Methodological framework for data completeness verification}, we propose a Frida-based dynamic monitoring framework, which can be divided into three main modules: identification of personal data categories, analysis of app’s personal data collection behavior and data comparison. Initially, because the range of personal data is too broad, we select ten data categories to monitor. Next, we select relevant sensitive data APIs for dynamic data monitoring based on these ten data categories. Finally, we compare the data copy obtained from the app with the data we actually collected to verify the completeness of the data copy.

\textbf{Identification of Personal Data Categories}:
In Section \ref{chap:two_first}, we give the legal and official definitions of personal information. In general, it includes all kinds of information related to identifiable natural persons. However, since the scope of user personal data is quite broad, we sample ten types of personal data (Table \ref{tab:Categories of Sensitive Personal Data}) from the three examples of personal information given by the CNSA: Network identity identification information; Personal frequently used device information; Personal location
information. While these ten categories of information do not cover all sensitive information about the user, by examining the data feedback from the application regarding these ten categories, we have been able to verify the completeness of the data copies.

\begin{table}[h]
  \centering
  \caption{Categories of Sensitive Personal Data}
  \label{tab:Categories of Sensitive Personal Data}
  \begin{tabular}{c}
    \toprule
    Data Category\\
    \midrule
    IP Address, Net Type, SSID, Android ID, OAID, AAID,\\ VAID, MCC/MNC, SIM Country Code, Location\\
  \bottomrule
\end{tabular}
\end{table}

\textbf{Analysis of App's Personal Data Collection Behavior}: To collect users' personal data, apps typically call certain Android native APIs, which can obtain users' sensitive data. Therefore, the first step in monitoring an app's data collection behavior is to gather these sensitive APIs. After reviewing the Android API reference \cite{android}, we identify 26 sensitive APIs covering the ten sensitive personal data categories in Table \ref{tab:Categories of Sensitive Personal Data}.

Based on these APIs, we aim to monitor the app dynamically. First, we write a JavaScript script responsible for capturing information and inject it into the target app using Frida. The script inserts code snippets into the app while it's running to monitor and intercept the app's calls to sensitive APIs. In this way, we can capture in real-time the app's interactions with these sensitive APIs, including the parameters used, the call stack, the data categories and the returned values. We also define a custom message callback function, which saves the sensitive API-related information from the JavaScript hooks, including the time of the API call, the operations performed, and the return value information. This data is stored in an CSV file for further data comparison and result analysis.

\textbf{Data Comparison}: Finally, we compare the data captured during runtime from the app with the data copy obtained from the app, verifying the completeness of the data copy.

When exercising RADS, we find that the data copy provided by apps can be in various formats, including CSV, JSON, HTML, TXT, and PDF. Most of the data copies are in CSV and JSON formats, which are relatively structured. Since other formats are less and their data formats are heterogeneous, we mainly perform automated analysis on CSV and JSON data copies, with manual comparison for other formats.

\textit{CSV Data Copy}: There are usually two data formats. One has the first row of each column as the data description (e.g., IP Address, Location), followed by specific data in each row; the other has the first column as the data description, with the second column as the specific data. Based on these two formats, we use the data descriptions as keys and the corresponding specific data as values to store in a dictionary.

\textit{JSON Data Copy}: JSON data is typically stored in a dictionary format, but it might contain nested structures where a key's value is a list of multiple dictionaries. To address this, we perform data flattening, concatenating nested keys, and building a new dictionary to avoid nested dictionaries in the values.

After processing, we obtain a dictionary with data descriptions and their corresponding specific data. We then summarize possible sensitive personal data descriptions in the data copy, for example, the possible descriptions of location information in a data copy are ``Location", ``Longitude" and ``Latitude". We use regular expressions to match data copy descriptions with previously defined sensitive personal data categories to extract specific sensitive personal data from the data copy.

Finally, we compare the extracted sensitive personal data from the data copy with the actual sensitive personal data collected during app runtime to verify the completeness of the data copy.

\section{Evaluation}
In this section, we will analyze the apps in G and H to answer the three research questions in Section \ref{chap:two_two}.
\subsection{Experiments for Answering RQ1}
In this part, we will identify RADS declarations and implementation methods in app privacy policies.

\textbf{Data Preparation}: In order to ensure the popularity and activity of the research applications, we select the mainstream application market G in Europe and the mainstream application market H in China. As of 2022, the number of monthly active users in these two markets exceeded 500 million, and G exceeded 2.5 billion. Since the official website of application market G does not list all its applications, but each category will list about dozens of applications with high popularity in the region where the IP is located, we obtain the privacy policies of 600 European applications in different categories in application market G. The privacy policies of applications in application market H are more convenient to obtain. We select applications that ranked high in different application categories, and finally obtain the privacy policies of 1031 applications in H, constructing our experimental dataset. It's worth noting that the privacy policies from apps in H are in Chinese, while our paragraph extraction work is based on English. Therefore, we use GPT-3.5 Turbo to translate these Chinese texts into English. 

\textbf{GPT-4-Based Identification Accuracy Evaluation}: To test the accuracy of our approach in Section \ref{chap:three_first}, we construct a test dataset. However, since there is currently no public dataset on RADS declarations and implementation methods, we build a test dataset using privacy policies from 50 apps randomly selected in each of the two app markets. Two classmates from our research group and I then manually identify, label, and verify the types and implementation methods of RADS. Using the identification approach, we identify the types and implementation methods of RADS declared in the privacy policies in our test dataset, and compare the results of our identification with those that had been manually annotated to evaluate the correctness of our approach. For each type and implementation method, we categorize applications into two groups, taking the right type VDAR as an example, we divide the applications into those that correctly identify VDAR and those that don't, and then assess them using binary classification evaluation metrics, including Accuracy, Precision, Recall, and the F1-Score. 

Table \ref{tab:Test Result for RADS and Implementation Method} summarizes our test results for the two categories of rights and the three implementation methods, where Acc represents Accuracy, P represents Precision, R represents Recall, and F1 represents F1-Score. In app markets G and H, for VDAR, our approach achieves 92\% and 98\% accuracy, with F1-Scores of 95.98\% and 98.95\%. For DCAR, our approach achieves 92\% and 92\% accuracy, with F1-Scores of 95.34\% and 94.59\%. Finally, for the three implementation methods of email contact, account settings and webform submission, our accuracy in app market G is 90\%, 94\% and 98\%, with F1-Scores of 93.32\%, 94.13\% and 94.74\%, respectively. And in app market H, our accuracy is 92\%, 90\% and 100\%, and F1-Scores of 92.59\%, 94.38\% and 100\%.

As can be seen from the results in the test set, our approach with the prompt designed for RADS is effective in identifying RADS and implementation methods declared in the privacy policy.

\begin{table}[h]
  \centering
  \caption{Accuracy Evaluation for Identification of RADS Declarations and Implementation Methods}
  \label{tab:Test Result for RADS and Implementation Method}
  \begin{tabular}{ccccccc}
    \toprule
    Market & & RADS & & Method & & \\
     & & VDAR & DCAR & Email & Settings & Webform\\
    \midrule
    G & Acc & 92.00\% & 92.00\% & 90.00\% & 94.00\% & 98.00\%\\
     & P & 97.83\% & 97.50\% &	94.59\% & 92.31\% &	90.00\%\\
     & R & 93.75\% & 92.86\% &	92.11\% & 96.00\% &	100\%\\
     & F1 & 95.98\% & 95.34\% &	93.32\% & 94.13\% &	94.74\%\\
    \midrule
    H & Acc & 98.00\% & 92.00\% & 92.00\% & 90.00\% & 100\%\\
     & P & 100\% & 100\% &	92.59\% & 100\% &	100\%\\
     & R & 97.92\% & 89.74\% & 92.59\% & 89.36\% &	100\%\\
     & F1 & 98.95\% & 94.59\% &92.59\% &94.38\% &	100\%\\
  \bottomrule
\end{tabular}
\end{table}

\textbf{Statistical Result Analysis}: Next, we automatically analyze the privacy policies of apps from two markets. As shown in Table \ref{tab:Proportion Statistics for RADS in G and H}, we calculate the proportion of three types of apps in two major markets: apps that declare VDAR, those that declare DCAR, and those that do not declare any about RADS. Firstly, in market G, apps that declare VDAR account for 14.17\%(85/600) of the total number of apps, while those that declare DCAR account for 54.50\% (327/600). This indicates that more than half of the apps in G fully comply with the GDPR's RADS regulation, providing DCAR. However, 31.33\% (188/600) of apps do not have any declaration about RADS. Secondly, in market H, the proportion of apps that fully comply with the GDPR's RADS regulation providing DCAR, is lower than in market G, with only 37.05\% (382/1031). Meanwhile, the proportion of apps that do not declare RADS is similar to that in market G, at 32.59\% (336/1031). This shows that there is a considerable compliance issue with RADS, over half of the apps needing improvement in terms of RADS compliance.

\begin{table}[h]
  \centering
  \caption{Proportion Statistics for RADS Declarations}
  \label{tab:Proportion Statistics for RADS in G and H}
  \begin{tabular}{ccc}
    \toprule
    RADS &G &H \\
    \midrule
    VDAR &14.17\%(85/600) &30.36\%(313/1031)\\
    DCAR &54.50\%(327/600) &37.05\%(382/1031)\\
    Non &31.33\%(188/600) &32.59\%(336/1031)\\
  \bottomrule
\end{tabular}
\end{table}

\begin{table}[h]
  \centering
  \caption{Proportion Statistics for Implementation Methods}
  \label{tab:Proportion Statistics for Implementation Method in G and H}
  \begin{tabular}{ccc}
    \toprule
    Method &G &H \\
    \midrule
    Email Contact &68.95\%(282/409)& 61.44\%(427/695)\\
    Account Settings &64.55\%(264/409)& 70.36\%(489/695)\\
    Webform Submission &19.07\%(78/409) &5.18\%(36/695)\\
  \bottomrule
\end{tabular}
\end{table}

Finally, we analyze the situation of implementation methods. As shown in Table \ref{tab:Proportion Statistics for Implementation Method in G and H}, there are two primary methods for exercising RADS: email contact and account settings, each accounting for around 60\%-70\%. Additionally, the method of webform submission is used more in G, at 19.07\% (78/409), while in H, it only accounts for 5.18\% (36/695). Overall, among the apps that provide RADS, about 64.22\% (709/1104) offer the email contact method, about 68.21\% (753/1104) offer the account settings method, and a small number of apps offer webform submission, about 10.33\% (114/1104).

\begin{flushleft}
\fcolorbox{black}{gray!10}{\parbox{0.98\linewidth}{\textbf{Answer to RQ1}: In market G, about 54.50\% of apps fully comply by declaring DCAR in their privacy policies, while about 31.33\% do not declare RADS. Similarly, in market H, only about 37.05\% of apps correctly declare DCAR, with about 32.59\% not declaring RADS. Additionally, the most common methods for exercising RADS are email contact and account settings, each accounting for around 60\%-70\%, while only 10.33\% of apps offer the webform submission.}}
\end{flushleft}

\subsection{Experiments for Answering RQ2}
In this part, we will evaluate the practicality of the implementation methods, focusing mainly on two aspects: authenticity and usability.

\textbf{Experiment Preparation}: We conduct a practicality evaluation of implementation methods on 200 top-ranked apps each in G and H, which can complete the registration process. The number of apps corresponding to methods of each data access right for each app market is summarized in Table \ref{tab:the Number of apps Corresponding to Methods of Each Data Access Right in G and H}. The reason for selecting a total of 400 apps instead of conducting practicality experiments on all 1,631 apps from the previous experiment is that the majority of these applications require manual methods such as facial recognition, mobile verification codes, and email authentication when executing RADS, which cannot be fully automated.

\begin{table}[h!]
    \centering
  \caption{Number of Apps Corresponding to Implementation Methods for RADS}
  \label{tab:the Number of apps Corresponding to Methods of Each Data Access Right in G and H}
  \begin{tabular}{ccccc}
    \toprule
    Method &G & &H & \\
     & VDAR &DCAR &VDAR &DCAR\\
    \midrule
    Email Contact &35 &72 &7 &51\\
    Account Settings &69 &30 &136 &60\\
    Webform Submission &12 &15 &0 &1\\
  \bottomrule
\end{tabular}
\end{table}

\textbf{Authenticity Analysis}: We first analyze the authenticity of the implementation methods. First of all, 17\% (34/200) of apps in G could genuinely provide users with their personal data copies, while in H, this proportion is 19\% (38/200). Next, we perform statistics on the authenticity of the implementation methods for each data access right in G and H, as shown in Table \ref{tab:Statistical Results of the Authenticity of Implementation Methods for Each Data Access Right in G and H}.
\begin{table*}
    \centering
  \caption{Statistical Results of the Authenticity of Implementation Methods for RADS}
  \label{tab:Statistical Results of the Authenticity of Implementation Methods for Each Data Access Right in G and H}
  \begin{tabular}{cccccc}
    \toprule
      & & G & &H & \\
    Method & Authenticity &Number &Proportion &Number &Proportion\\
    \midrule
    Email Contact &Failure &81 &75.70\% &47 &81.03\%\\
     & View Information &8 &7.48\% &7 &12.07\%\\
     & Obtain Data Copy &18 &16.82\% &4 &6.90\%\\
    \midrule
    Account Settings &Failure &12 &12.12\% &3 &1.53\%\\
     & View Information &64 &64.65\% &156 &79.59\%\\
     & Obtain Data Copy &23 &23.23\% &37 &18.88\%\\
    \midrule
    Webform Submission &Failure &19 &70.37\% &1 &100\%\\
     & View Information &0 &0\% &0 &0\%\\
     & Obtain Data Copy &8 &29.63\% &0 &0\%\\
  \bottomrule
\end{tabular}
\end{table*}


From an overall perspective, among apps that offer account settings method, about 20.34\% (60/295) could actually provide data copies. Similarly, among those that offer email contact method, the proportion is 13.33\% (22/165), while the proportion reaches 28.57\% (8/28) for apps offering webform submission method. It is worth mentioning that the total number of applications corresponding to the above three types of execution methods is greater than the total number of applications providing copies in the two application markets mentioned above. This is because we may obtain the data copy through multiple methods in an application. Finally, among those that can not provide data copies, some apps offer a weaker user experience by allowing users to view personal information online within the app.

Based on this, most apps that provide account settings method allow users to view personal information through this method. In market G, about 12.12\% (12/99) of apps that declare the account settings method could not view personal information, while in market H, only about 1.53\% (3/196). This small portion might have outdated privacy policies, causing the specified implementation method in the policy to be incorrectly executed in the mobile app, leading to an inability to obtain relevant user information. However, the majority of apps that offer email contact as an implementation method could not view personal information through this method, with about 75.70\% (81/107) in G and 81.03\% (47/58) in H. This is mainly due to apps failing to respond after we email to them. A small number of apps, while unable to provide data copy, offer specific steps for viewing personal information within the app in their responses. Similarly, most apps that offer webform submission could not view personal information through this method, with about 70.37\% (19/27) in G. H had too few samples, with only one app offering webform submission, which also did not provide any feedback.

Moreover, among apps that provide DCAR and account settings method, about 76.67\% (23/30) in market G could genuinely offer users a personal data copy, while in H, this proportion is about 61.67\% (37/60). For email contact, the proportion of apps in market G that could genuinely obtain a data copy is 25\% (18/72), while in market H, it is about 7.84\% (4/51). Similarly, for webform submission, in G, about 53.33\% (8/15) could obtain a data copy, while in H, it is 0\% (0/1).

\begin{figure}[h!]
  \centering
  \includegraphics[width=\linewidth]{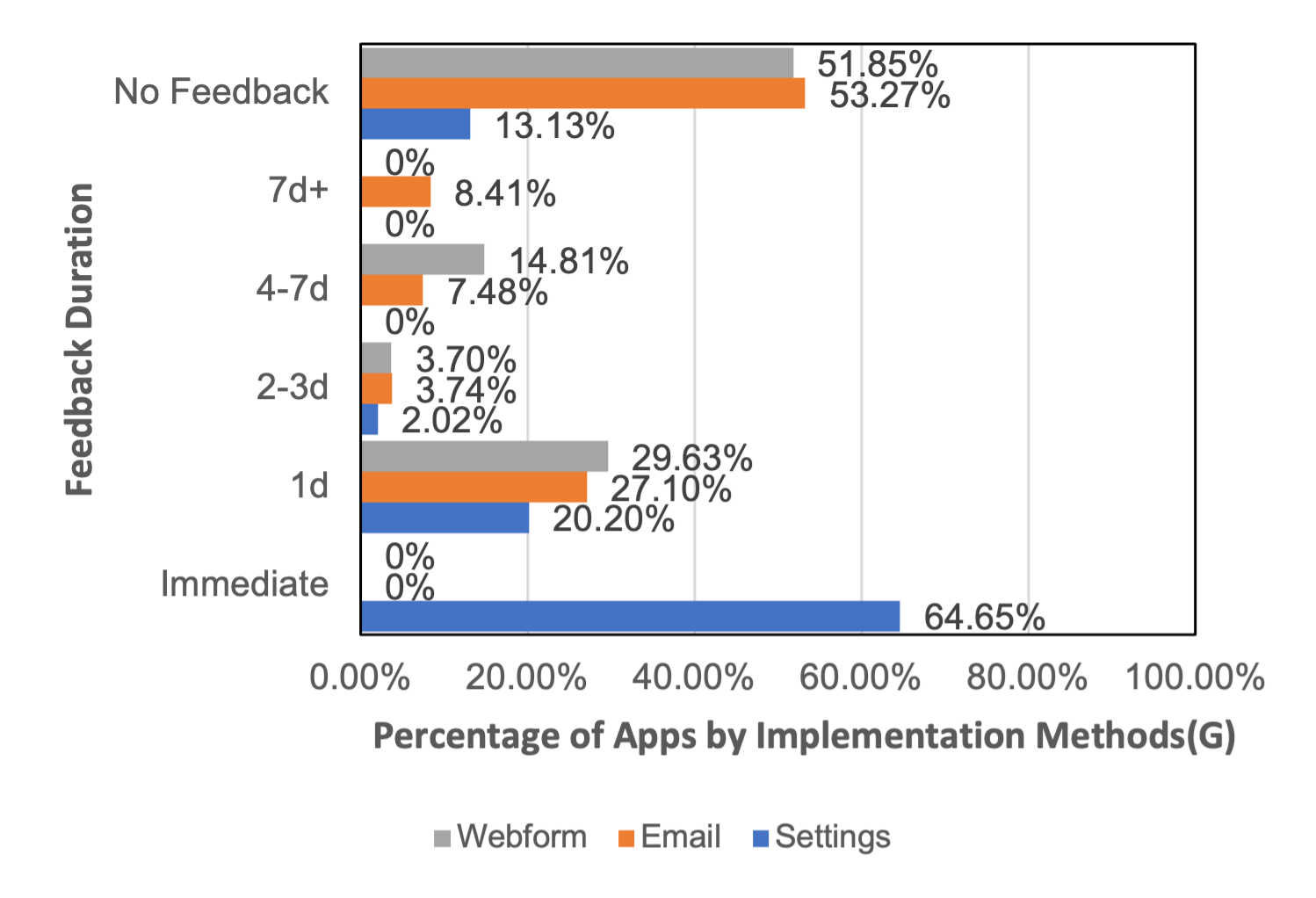}
  \caption{Statistical Results on Feedback Duration of apps in G}
  \label{fig:Statistics on feedback duration of apps in G}
\end{figure}

\begin{figure}[h!]
  \centering
  \includegraphics[width=\linewidth]{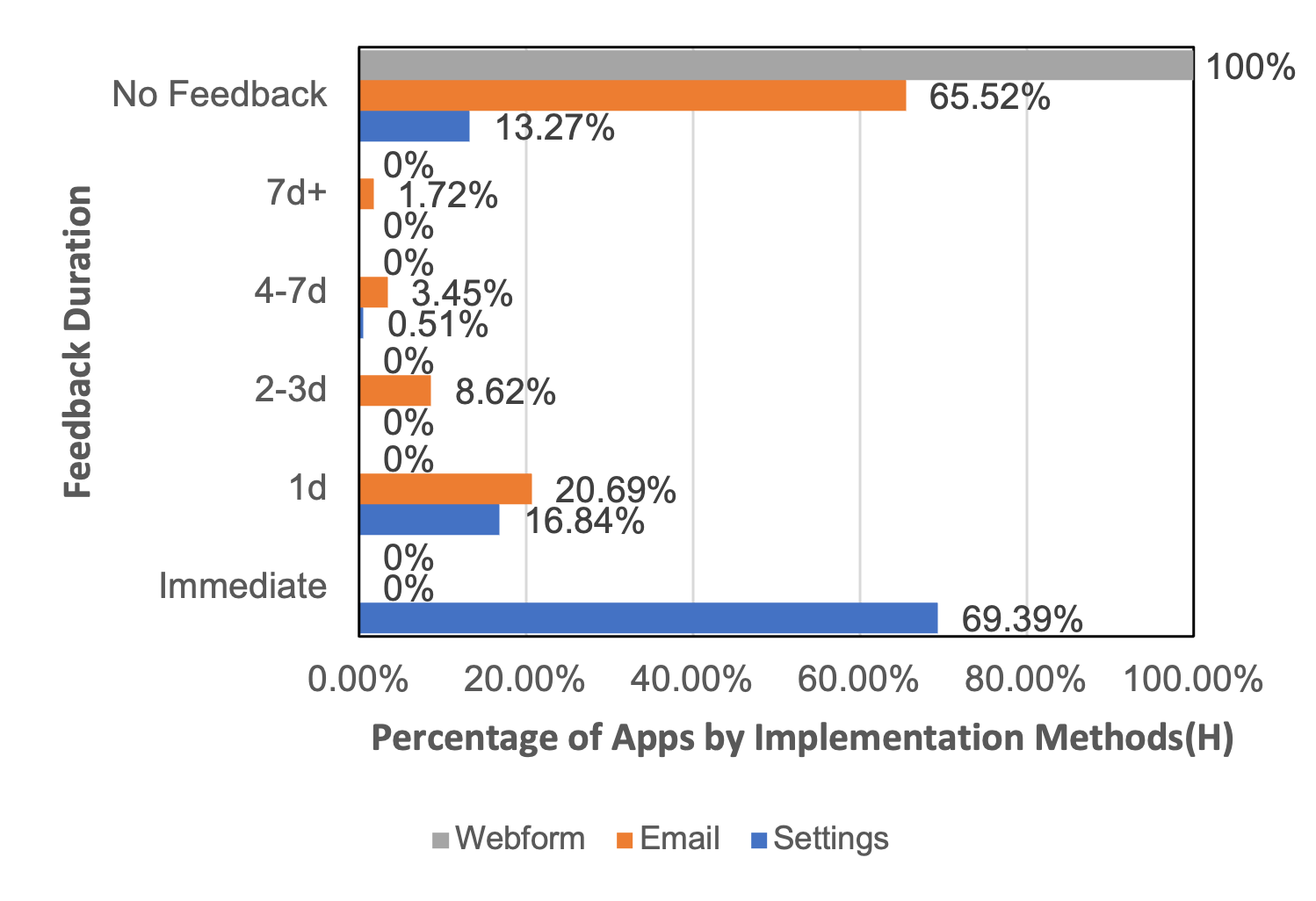}
  \caption{Statistical Results on Feedback Duration of apps in H}
  \label{fig:Statistics on feedback duration of apps in H}
\end{figure}

\textbf{Usability Analysis}: Firstly, we focus on the feedback duration for each implementation method to analyze the usability, as shown in Figure \ref{fig:Statistics on feedback duration of apps in G} and Figure \ref{fig:Statistics on feedback duration of apps in H}. We divide feedback duration into six categories: immediately viewing, 1 day, 2-3 days, 4-7 days, over 7 days, and no feedback. Immediately viewing is a special category for account settings, where users can view personal information online within the app settings.

\textit{Account Settings}: In this method, the proportion of apps that allow immediately viewing is between 60\% and 70\% in G and H. In the apps where a request for a data copy is submitted through account settings, about 13\% don't provide feedback in both markets. Among those that provide feedback, the majority respond within one day, accounting for 20.20\% (20/99) in G and 16.84\% (33/196) in H. Only 2.02\% (2/99) of apps in market G have feedback duration of 2-3 days.

\textit{Email Contact}: The proportion of apps that don't provide feedback increases significantly, reaching 57.58\% (95/165). In G, it accounts for 53.27\% (57/107), and in H, it accounts for 65.52\% (38/58). Additionally, unlike account settings, the feedback duration for email contact is lengthened. While apps with a feedback duration within one day are still the majority, with about 27.10\% (29/107) in G and about 20.69\% (12/58) in H, there are also apps with feedback duration ranging from 4-7 days or more than 7 days. Apps with a feedback duration of over 4 days account for 15.89\% (17/107) in G and 5.17\% (3/58) in H. Overall, among the apps that provide feedback, the proportion of those with feedback duration exceeding 4 days is 28.57\% (20/70). Apps with a feedback duration of over 7 days account for 6.06\% (10/165), but among those that provide feedback, it is 14.29\% (10/70).

\textit{Webform Submission}: Similar to email contact, about 53.57\% (15/28) of apps don't provide feedback, of which apps in G account for 51.85\% (14/27) and H account for 100\% (1/1). The majority of apps that provide feedback have a feedback duration of within one day, accounting for about 29.63\% (8/27). The second most common feedback duration is 4-7 days, with a proportion of about 14.81\% (4/27). The 2-3 day feedback duration is the least common, with a proportion of about 3.81\% (1/27). No apps are found with a feedback duration exceeding 7 days.

\begin{figure}[h!]
  \centering
  \includegraphics[width=\linewidth]{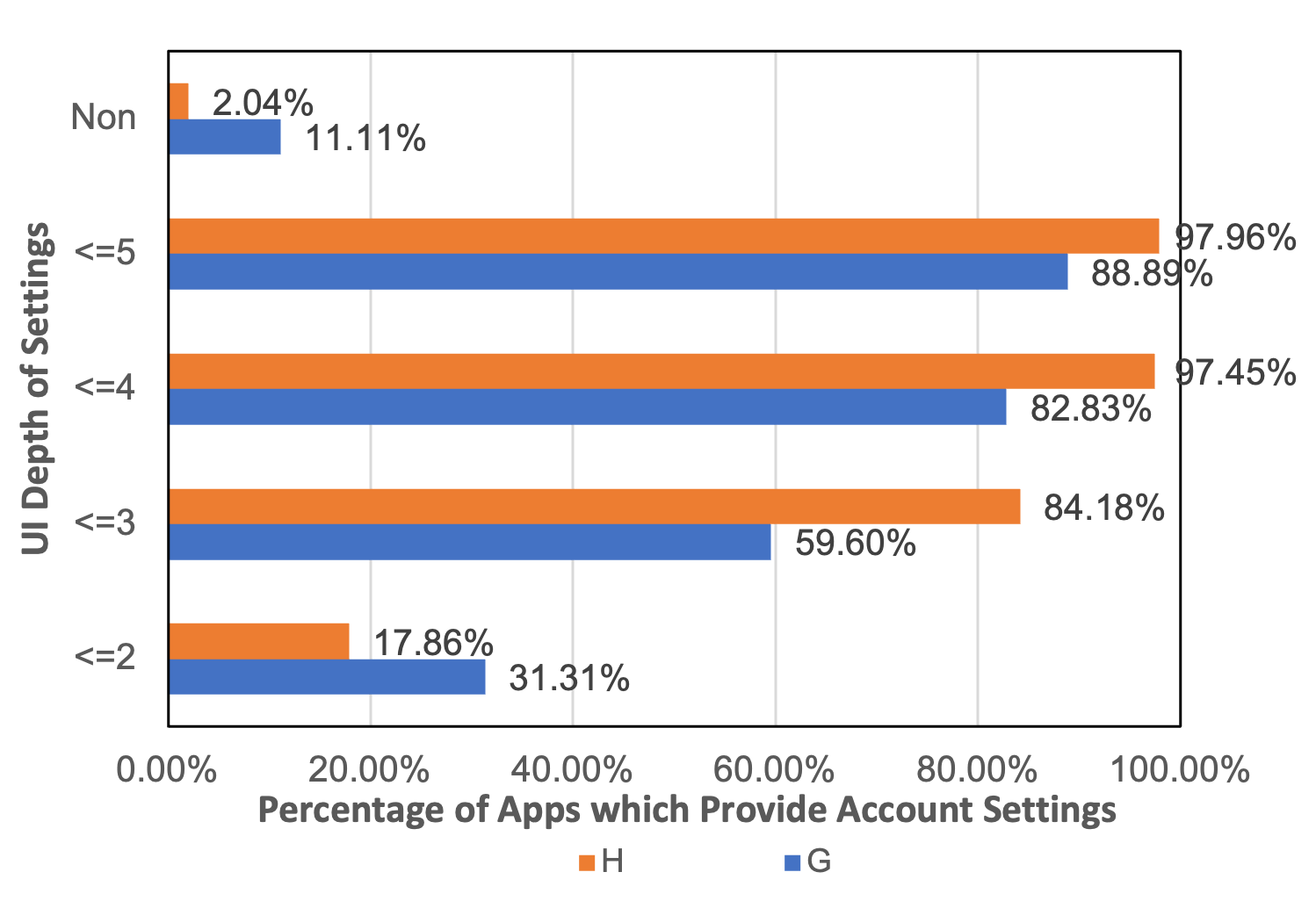}
  \caption{Statistical Results on UI Depth of apps}
  \label{fig:Statistics on Settings Depth of apps in G and H}
\end{figure}

Finally, we analyze the UI depth of settings, as shown in Figure \ref{fig:Statistics on Settings Depth of apps in G and H}. Our practice finds that the UI depth ranges from 2 to 5 levels. In both markets, the majority of apps have a UI depth of 4 levels or less, accounting for 82.83\% (82/99) in G and reaching 97.45\% (191/196) in H. Besides, the distribution of apps with UI depth of 2, 3 and 4 is relatively even in G, while in H, the majority of apps have an UI depth of 3 levels, accounting for 66.33\% (130/196). Additionally, some account settings declared in the privacy policy couldn't be found, accounting for 11.11\% (11/99) in G and 2.04\% (4/196) in H.

\begin{flushleft}
\fcolorbox{black}{gray!10}{\parbox{0.96\linewidth}{\textbf{Answer to RQ2}: Overall, the proportions of apps that can truly provide data copies are 17\% (34/200) in G and 19\% (38/200) in H. Additionally, the performance of all three methods for obtaining data copies is poor, with each accounting for less than 30\%. The feedback duration for account settings is mostly within one day, but the other two methods show worse performance. For email contact, 57.58\% (95/165) of apps do not provide feedback, while among those that do, about 28.57\% (20/70) have a feedback duration of over 4 days, and about 14.29\% (10/70) over 7 days. Similarly, for webform submission, 53.57\% (15/28) of apps do not provide feedback. Finally, about 93\% of apps have an UI depth of 4 levels or less for settings, with around 5\% of apps having settings that do not align with the description in the privacy policy.}}
\end{flushleft}

\subsection{Experiments for Answering RQ3}
In this part, we analyze the consistency between the data copy obtained by exercising RADS and the actual data collected by the app, thereby verifying the completeness of the data copy.

\begin{table*}
    \centering
  \caption{Comparison Results of Different Data Categories' Consistency among apps in G and H}
  \label{tab:Comparison Results of Different Data Categories' Consistency among apps in G and H}
  \begin{tabular}{ccccccccccc}
    \toprule
     Market &IP Address	&Net Type&SSID&Android ID	&OAID	&AAID	&VAID	&MCC/MNC	&SIM Country Code	&Location \\
    \midrule
    G	&48.15\%	&2.94\%	&0\%	&30.77\%	&0\%	&0\%	&0\%	&28.57\%	&33.33\%	&42.86\%\\
    H	&0\%	&0\%	&0\%	&0\%	&0\%	&0\%	&0\%	&88.57\%	&89.47\%	&13.33\%\\

  \bottomrule
\end{tabular}
\end{table*}

\textbf{Experiment Preparation}: In the previous section, we performed RADS on a total of 400 apps. After exercising RADS, we obtain a total of 72 data copies, of which 34 are from apps in G, and 38 are from apps in H. For these 72 apps, we use the approach from Section \ref{chap:three_third} to dynamically monitor the behavior of the apps and analyze the actual personal data collected by the apps during runtime. For data copies in CSV and JSON formats, we extract the sensitive data of interest through an automated approach, while for other formats, we manually extract the data.

\textbf{Completeness Analysis}: Table \ref{tab:Comparison Results of Different Data Categories' Consistency among apps in G and H} shows the consistency comparison results for different data categories in both markets. We find that only a small portion of apps collect SSID, AAID, and VAID, while most apps collect the other seven categories of data. Overall, apps perform better in the MCC/MNC and SIM Country Code categories, with 61.90\% (39/63) and 62.16\% (23/37) of apps providing these two categories of data correctly. However, the performance is poor for other data categories. In the IP Address, Network Type, and Location, the proportions of apps providing the corresponding data correctly are 22.81\% (13/57), 1.43\% (1/70), and 27.59\% (8/29), respectively. Not a single app correctly provide information for the other four data categories.

For IP Address, 48.15\% (13/27) of apps in G that collect user IP address provide IP correctly in the data copy. However, in H, while most apps collect the IP address, none provide the IP Address or the provided IP Address doesn't match the one actually collected.

For Network Type and Android ID, although most apps in both markets collect these two data categories, only about 3\% of apps in G provide the collected data in data copy. For OAID, although most apps in H collect OAID, none provide the correct data.

Regarding MCC/MNC and SIM Country Code, the performance between the two markets differs significantly, with about 30\% of apps in group G and about 89\% in group H correctly providing the collected data.

For Location, less than half of the apps in both markets provide correct Location information. Specifically, in G, 42.86\% (6/14) of apps provide correct location information, while only 13.33\% (2/15) of apps in H do so.

\begin{figure}[h!]
  \centering
  \includegraphics[width=\linewidth]{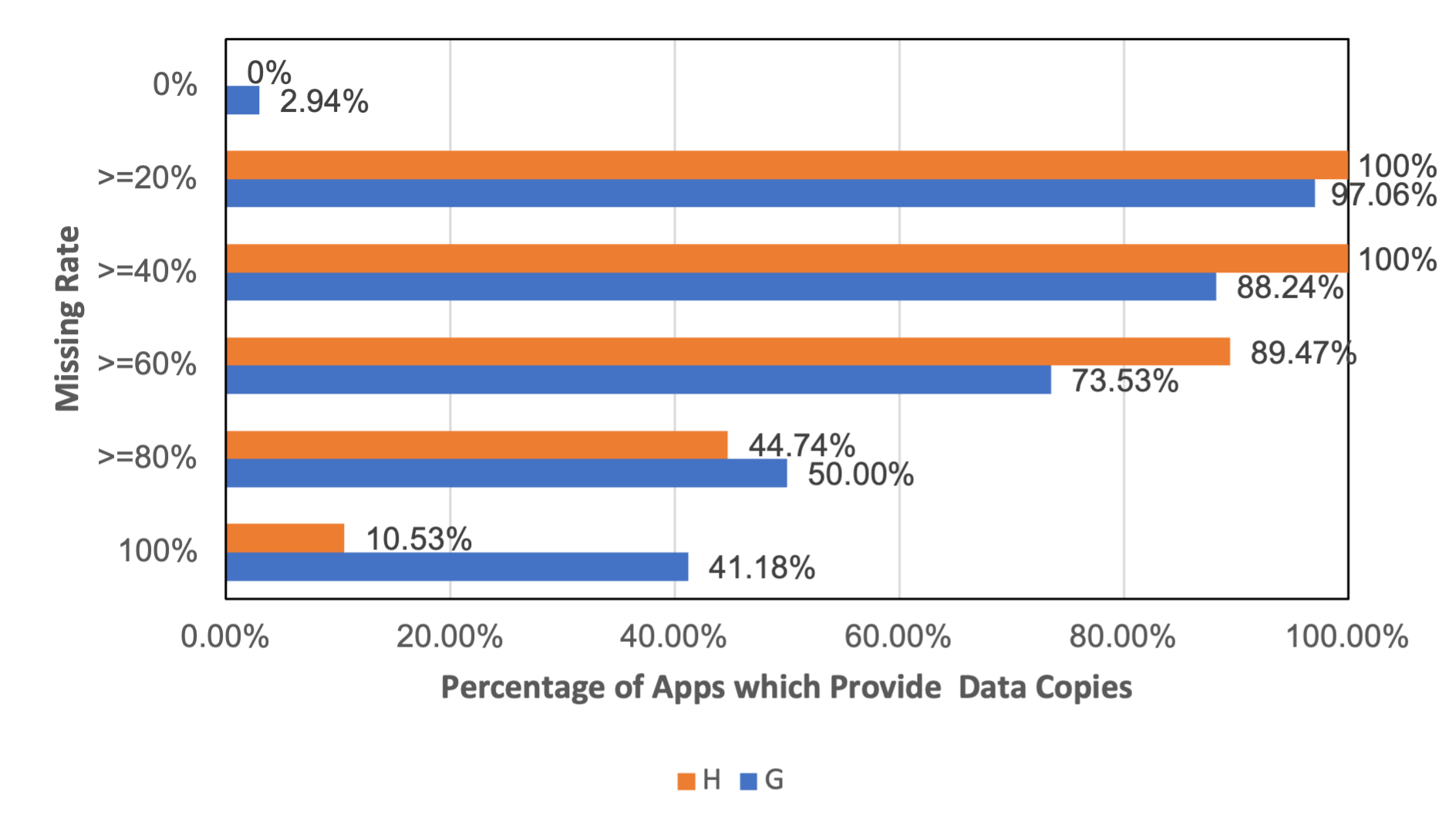}
  \caption{Statistical Results of Missing Rate in G and H}
  \label{fig:Statistical Result of Missing Rate in G and H}
\end{figure}

We further calculate the percentage of data categories that could not be successfully matched in the actual collected data (referred to as the missing rate), with the results shown in Figure \ref{fig:Statistical Result of Missing Rate in G and H}. We find that only about 2.94\% (1/34) of apps in G ensure the completeness of the data copy, with the rest having issues of missing involving personal data. Additionally, in both markets, nearly half of the data copies have the missing rate exceeding 80\%, with about 44.74\% (17/38) in H and about 50.00\% (17/34) in G. This indicates that the data copy rarely contains actual personal information related to users. Moreover, the vast majority of apps have an missing rate exceeding 40\%, with about 88.24\% (30/34) in G and 100\% (38/38) in H. It is clear that the data copy provided by apps currently cannot guarantee completeness to a large extent, indicating substantial room for improvement.

\begin{flushleft}
\fcolorbox{black}{gray!10}{\parbox{0.96\linewidth}{\textbf{Answer to RQ3}: Consistency analysis of data categories reveals that, aside from MCC/MNC and SIM Country Code, where about 62\% of apps could correctly provide personal data, other data categories have poor consistency performance. Notably, there are four data categories (SSID, OAID, AAID, VAID) in which no app could provide the corresponding information correctly. Additionally, from the perspective of data missing rates, only 2.94\% (1/34) of apps in G could provide a complete data copy, and the other apps fail to comply with legal regulations, leading to the missing of personal data. Furthermore, nearly half of the apps have an missing rate exceeding 80\%, and over 90\% of apps have an missing rate exceeding 40\%. This indicates that the completeness of the data copy is far from being ensured.}}
\end{flushleft}

\section{Related Work}
We divide related work into four categories: Right of Access by the Data Subject, privacy policy analysis, mobile app analysis, and GDPR compliance detection.

\textbf{Right of Access by the Data Subject}: Borem et al.\cite{borem2024reactions} study user responses to personal data obtained through Data Subject Access Requests and find that current data export content falls short of user needs. Martino et al.\cite{martino} conduct a longitudinal analysis of the authentication process in the GDPR access policy, and discover the privacy risks brought by the weak authentication policy. Lobel et al. \cite{darkpatterns2023access} study how popular websites use dark patterns to discourage users from trying to exercise RADS. Unlike previous work that mostly used user surveys, we mainly studied RADS through dynamic analysis and comparison.

\textbf{Privacy Policy Analysis}: Del Alamo et al. \cite{Del2022} recently conduct a systematic survey of automated privacy policy analysis techniques, most of which use codification or annotation methods (Saldana \cite{saldana2021coding}). This involves one or more domain analysts systematically labeling policy statements to create structured annotations of privacy practices (i.e., a corpus). Privee \cite{Privee2014} and Polisis \cite{Polisis2018} analyze privacy policies at the document and paragraph levels to answer users' questions, but they are limited by coarse granularity. Some useful corpora have been released in the privacy field (Zimmeck et al. \cite{Zimmeck2019MAPSSP}; Wilson et al. \cite{wilson2016}), which are used as benchmark data for building automated classification models. For example, Zimmeck et al. \cite{Zimmeck2019MAPSSP} automatically extract data collection practices from privacy policies. Additionally, PolicyLint \cite{Policylint2019} uses dependency parsing to extract privacy statements from policy documents but does not analyze the purpose of data collection. Bhatia et al. \cite{Bhatia2017} identify common patterns of purposive statements from privacy policies and use semantic frameworks to analyze privacy goals' incompleteness, including the purposes of data practices \cite{Bhatia2018}. Shvartzshnaider et al. \cite{Shvartzshnaider2019GoingAT} analyze the information flow in a limited set of privacy policies according to the contextual integrity framework \cite{Nissenbaum2004PrivacyAC}. Compared with previous work, our research uses the large language model GPT-4\cite{openai2024gpt4} to analyze privacy policies. GPT-4, a large pre-trained language model based on the Transformer architecture \cite{transformer},  was pre-trained on a massive text dataset and can capture deep semantic and structural features of language \cite{LLMSurvey}. GPT-4's excellent text understanding capabilities allow us to identify privacy terms in privacy policy texts, ensuring high recognition accuracy and precision \cite{LIU2023100017,wei2022emergent}.

\textbf{Mobile App Analysis}: Researchers mainly use static analysis, dynamic analysis, and hybrid analysis to examine mobile app behavior (Del Alamo et al. \cite{Del2021}). For example, Zimmeck et al. \cite{Zimmeck2016AutomatedAO} use the static analysis tool Androguard with PScout to check whether apps make sensitive Android API calls. Ferrara and Spoto \cite{ferrara2018} use static code analysis to detect personal data disclosure for data protection officers to identify potential GDPR violations. Jia et al. \cite{Jia2019} employ dynamic analysis techniques to detect personal data disclosure in network packets without user consent. Wang et al. \cite{GUILeak} develop a GUI-based method to automatically detect privacy leaks of user-input data in specific Android apps and determine whether such leaks could violate the app's privacy policy statements. The PoliCheck \cite{PoliCheck} tool defines an automated, entity-sensitive flow-based consistency analysis to detect whether mobile apps disclose the handling of privacy-sensitive data flows properly. Nguyen et al. \cite{nguyen} study violations in Android apps that send personal data to data controllers without prior explicit user consent. Given the limitations of static analysis, which can't capture the actual behavior of programs, our work dynamically analyzes user data collected by mobile apps during runtime based on the Frida framework.

\textbf{GDPR Compliance Detection}: Regarding GDPR compliance, Fan et al. \cite{Fan2020} conduct an empirical evaluation of Android mobile health apps, examining their transparency, data minimization and confidentiality requirements, focusing on six practices disclosed through privacy policies. Klein et al. \cite{Klein2023} treat data protection law compliance as an information flow tracking problem, designing a dynamic information flow tracking framework called Fontus to detect GDPR compliance during app data processing. Zhou et al. \cite{Zhou2023POLICYCOMPCC} propose an automated framework, POLICYCOMP, to detect overbroad personal data collection practices in privacy policies. Danny et al. \cite{Danny2023AutomatedGC} suggest a fully automated approach for assessing whether Android apps comply with GDPR requirements for cross-border personal data transfers. Compared to previous work, our research focuses on RADS specified in the GDPR.

\section{Discussion}
In this section, we conducted an in-depth analysis of the research results and current situation, and provided some suggestions for the existing issues.

Our research results indicate that despite the latest data protection regulations worldwide, such as GDPR and PIPL, declaring users' rights to access personal data, the actual practice of RADS in the current Android app market is not optimistic. Next, we analyze the research results and current situation from three aspects of the RADS-Checker framework and provide some suggestions for the existing issues.

\subsection{Analysis of Results and Current Situation}
Firstly, regarding the declaration of RADS in privacy policies, we found that only around 43.5\% of apps fully comply with declaring DCAR in their privacy policies, while approximately 32.1\% of apps do not have any declaration regarding RADS in their privacy policies. For these non-compliant privacy policies, we believe that on the one hand, some app developers may use Automated Privacy Policy Generators (APPGs) to automatically generate privacy policies for convenience or cost-saving purposes. Pan et al. \cite{Pan2023} found in their investigation that 20.1\% of privacy policies in Google Play are generated by existing APPGs. However, these APPGs themselves have certain design flaws, leading to inadequate declaration of data rights and other privacy non-compliance issues. On the other hand, some app developers subconsciously believe that users will not spend a significant amount of time reading and understanding the content of privacy policies, and most users do not have sufficient legal knowledge to identify the issues in privacy policies, so they often do not pay enough attention to privacy policies.

Secondly, by evaluating the practicality of the implementation methods, we found that only around 18\% of apps are able to provide data copies to users. From the perspective of feedback time, more than half of the apps have feedback times exceeding 7 days or do not provide any feedback. From the perspective of the depth of UI settings, around 93\% of apps  have an UI depth of 4 levels or less for settings. However, it is worth noting that since users are not aware of the specific location of the UI settings interface beforehand, they can only rely on their experience and intuition to click through the app interface. Therefore, the number of clicks required for users, from opening the app to finding the UI settings option, is much more than 4. Thus, the long feedback time and hidden UI settings significantly undermine the enthusiasm and success rate of users exercising RADS. We analyze that this situation is mainly due to the lack of strict regulations in the laws and regulations regarding these details, so app service providers, from their own convenience perspective, vaguely declare that they will provide feedback within a month or even longer, without guaranteeing the timeliness of processing data access requests.

Lastly, through completeness verification of data copies, we found that only 2.94\% (1/34) of apps in G could provide a complete data copy, and the other apps do not comply with legal regulations, leading to the missing of personal data. More than 90\% of apps have data omission rates exceeding 40\%. For this situation, we believe that the format, organization, and design of data copies vary greatly among different companies. The scope of data downloads can range from individual CSV files to massive archives containing hundreds of gigabytes of data \cite{Veys2021}. Many files use formats like JSON or CSV, which often utilize UNIX timestamps instead of human-readable dates and times. Furthermore, even more typical human-readable data formats like HTML can be disorganized, filled with jargon and terminology. Therefore, there are significant deficiencies in the transparency and usability of data copies. Additionally, our research results found that a portion of the data copies provided by apps only include the account information provided by users during registration and completely ignore the sensitive data generated by users while using the app, such as IP addresses and geographical locations. We believe that app service providers lack a comprehensive mechanism for handling users' privacy data, resulting in the inability to timely and accurately integrate all users' privacy data when processing data copy access requests, thus unable to provide complete data copies.

\subsection{Suggestions}
To address the aforementioned issues, we put forward some suggestions, in order to better standardize data practices and protect user data.

\textbf{Raise awareness.} Policy-makers and the media should enhance publicity to make the public aware of users' fundamental data right to access and obtain a copy of their personal data from the data controllers, and exercising RADS as an important way to protect user privacy. App developers and service providers should be urged to comply with laws and regulations, providing complete declarations of RADS and enforcement methods in privacy policies.

\textbf{Simplify data access operations.} App developers are advised to provide unified data access enforcement methods for account settings and place UI setting options in a location that is easy for users to locate, such as ``Settings -\textgreater Account Security -\textgreater Data Copies", so as to reduce the cost for users to find the option, and at the same time, provide timely feedback on users' data access requests, and limit the feedback time to less than 7 days.

\textbf{Improve data transparency.} Provided data copies should be meaningfully organized, filtered and aggregated by data type or other attributes to improve data readability, and furthermore, user-friendly interactive interfaces can be provided to improve data usability and transparency.

\section{Conclusion}
In this paper, we propose a compliance measurement framework for RADS in apps: RADS-Checker, and explore RADS compliance in the current app markets.

First, to analyze the compliance of current apps with legal requirements for RADS and corresponding implementation methods, we categorize the RADS declarations in app privacy policies into two categories: DCAR and VDAR, and also categorize the implementation methods similarly. We then proceed with paragraph extraction of privacy policy texts and further analyze the extracted paragraphs using the large language model GPT-4. We find that about 54.50\% of apps in G fully comply with DCAR in their privacy policies, while in H, the compliance rate is only 37.05\%.

Next, we analyze the practicality of implementation methods, including two parts: authenticity and usability. We find that the proportion of apps that can truly provide data copies is less than 20\% in both markets. Among the three methods, the account settings method shows good performance in terms of authenticity, with more than 90\% of apps that allow users to view personal information. However, the email contact and webform submission methods perform poorly, with a rate of less than 30\%. Additionally, the account settings method typically has a feedback duration within one day. However, the other two methods perform poorly. For the email contact, 57.58\% of apps do not provide feedback. Similarly, for the webform submission, 53.57\% of apps do not provide feedback.

Finally, we analyze the completeness of the data copy provided by the app. By dynamically monitoring apps during runtime based on sensitive user-related APIs and comparing the real data observed with the data copies returned to users, we find that only 2.94\% of apps in G could provide a complete data copy, while the vast majority of apps do not comply with laws to ensure RADS. These findings suggest that there is considerable room for improvement in the compliance of the app with RADS.

\bibliographystyle{IEEEtran}
\bibliography{citepaper}


\newpage

 





\end{document}